\documentclass[10pt,conference]{IEEEtran}

\IEEEoverridecommandlockouts                              % This command is only needed if you want to use the \thanks command
\usepackage{cite}
\usepackage{amsfonts,amsmath,color,amssymb,amsxtra,amsbsy,floatflt}
\usepackage{graphicx}
\usepackage{algorithmic}
\usepackage{textcomp}
\usepackage{xcolor}
\usepackage{caption}
\usepackage{float}
\usepackage{comment}
\usepackage{url}
\usepackage{booktabs}
\usepackage{microtype}
\usepackage[caption=false,font=footnotesize]{subfig}

\def\BibTeX{{\rm B\kern-.05em{\sc i\kern-.025em b}\kern-.08em
    T\kern-.1667em\lower.7ex\hbox{E}\kern-.125emX}}

\newcommand{\new}[1]{{\color{black}{#1}}}

\title{\LARGE \bf Network Digital Twin for 5G-Enabled Mobile Robots}

\author{\IEEEauthorblockN{
Luis Roda-Sanchez\IEEEauthorrefmark{1},
Lanfranco~Zanzi\IEEEauthorrefmark{2},
Xi Li\IEEEauthorrefmark{2},
Guillem Gar\'i\IEEEauthorrefmark{3},
Xavier~Costa-P\'erez\IEEEauthorrefmark{4}\IEEEauthorrefmark{2}\IEEEauthorrefmark{5}}
\IEEEauthorblockA{
\IEEEauthorrefmark{1} Universidad de Castilla-La Mancha, Albacete, Spain,
Email:luis.roda@uclm.es\\
\IEEEauthorrefmark{2} NEC Laboratories Europe, Heidelberg, Germany, Email:\{name.surname\}@neclab.eu,\\
\IEEEauthorrefmark{3} Robotnik, Valencia, Spain.
Email:ggari@robotnik.es}
\IEEEauthorrefmark{4} i2CAT Foundation, Barcelona, Spain.
\IEEEauthorrefmark{5} ICREA, Barcelona, Spain.
}

\renewcommand{\baselinestretch}{0.975}

\begin{document}

\maketitle
\thispagestyle{empty}
\pagestyle{empty}

%%%%%%%%%%%%%%%%%%%%%%%%%%%%%%%%%%%%%%%%%%%%%%%%%%%%%%%%%%%%%%%%%%%%%%%%%%%%%%%%
\begin{abstract}
%The integration of robotic systems into exploration processes paved the road for a transformative synergy between technology and the environment.
The maturity and commercial roll-out of 5G networks and its deployment for private networks makes 5G a key enabler for various vertical industries and applications, including robotics. Providing ultra-low latency, high data rates, and ubiquitous coverage and wireless connectivity, 5G fully unlocks the potential of robot autonomy and boosts emerging robotic applications, particularly in the domain of autonomous mobile robots. %To boost the full potential of these robotic systems, it is of key importance to gain a comprehensive understanding of their surrounding network infrastructure.
Ensuring seamless, efficient, and reliable navigation and operation of robots within a 5G network requires a clear understanding of the expected network quality in the deployment environment. However, obtaining real-time insights into network conditions, particularly in highly dynamic environments, presents a significant and practical challenge.
In this paper, we present a novel framework for building a Network Digital Twin (NDT) using real-time data collected by robots. This framework provides a comprehensive solution for monitoring, controlling, and optimizing robotic operations in dynamic network environments.
We develop a pipeline integrating robotic data into the NDT, demonstrating its evolution with real-world robotic traces. We evaluate its performances in radio-aware navigation use case, highlighting its potential to enhance energy efficiency and reliability for 5G-enabled robotic operations.
%We further develop a pipeline that integrates robotic data into the NDT, demonstrating how the NDT evolves within an operational setting using real-world robotic traces and data. To show the applications of NDT, we give an example of using it for radio-aware navigation and highlight its potential benefits in improving energy saving and reliability for 5G robot operations.    
%NDT acts as a predictive tool by leveraging advanced analytic and machine learning algorithms. It anticipates potential network bottlenecks or failures, allowing for preemptive measures to maintain uninterrupted robotic operations.
\end{abstract}
%%%%%%%%%%%%%%%%%%%%%%%%%%%%%%%%%%%%%%%%%%%%%%%%%%%%%%%%%%%%%%%%%%%%%%%%%%%%%%%%

\section{Introduction}
\label{sec:intro}

% Over the past few years, there has been notable progress in wireless communication technologies, with 5G networks emerging as a crucial enabler for a wide range of applications. Simultaneously, 
The field of robotics has experienced notable progress in the realm of autonomous mobile robots, now equipped with wireless technologies like WiFi and 5G, enabling remote operation and task offloading~\cite{DT_survey}. These robots hold the promise of transforming various industries by autonomously and efficiently performing tasks in dynamic environments. 

% Xi: added this paragraph to highlight the the motivation of this work,  why we need an NDT for robots and what is the gap.
%
Autonomous robots rely on onboard sensors to build environmental knowledge of their proximity by collecting sensor data along their navigation trajectory, processing this information to build a comprehensive understanding of their environment~\cite{SLAM}.
% Nevertheless, for seamless, efficient, and reliable navigation and operation, these robots must not only have a comprehensive understanding of their physical surroundings (encompassing accurate perception and mapping) but also require knowledge of the networks they are connected to as to guarantee reliable communications. 
This results in expensive and time-consuming navigation efforts when a global network quality of the whole environment is needed. At the same time, varying radio quality can hinder the data acquisition process, 
% leading to increased latency and reduced localization accuracy, 
which in turn affects the robot's ability to make timely decisions and maintain precise positioning.
To tackle this challenge, Network Digital Twin (NDT) technology emerges as a promising solution, offering the ability to estimate the expected quality of a physical network. It does so by leveraging a combination of measured or estimated data collected from both the network infrastructure and the robots or other User Equipment (UEs).
Such NDT is not merely a static model but a dynamic and data-driven one that mirrors the behaviour, components, and interactions of the physical network in near real-time~\cite{next_generation_DT_robotics}. %Such technological innovation has the potential to revolutionize how we design, plan, optimize, and manage complex networks, ultimately aiming to enhance performance, reliability, cost savings, and sustainability.
This would allow mobile robots to make more informed decisions and adapt their autonomous operations over run time. For instance, mobile robots would benefit from optimal radio link conditions for computational offloading, helping conserve energy that would otherwise be used to transmit data under poor channel conditions.
%The Digital Twin (DT) technology has been adopted in many fields like industry, energy, transportation, and health care areas, 
However, the research of NDT technology in the field of communication networks is still in its early phase. Most of the NDT works rely on collected or simulated network data, and take a system-wide approach from a network operator's point of view by assuming full knowledge of the network infrastructure. The drawback of such approach is not only costly (in terms of model complexity and large amount of data to collect), but also slow to adapt to network changes, which is incapable of capturing dynamic changes over run time.  Conversely, we aim to build an NDT from the UE (or a robot) perspective, which is not aware of the deployment environment and of the available network infrastructure. 
% Besides, how to leverage robotic data to construct an NDT is new in the field. 

Thus, in this paper we propose a novel framework to build an online NDT from a robot when exploring an unknown area and its corresponding network environment through Robot Operating System (ROS), for estimation of the radio quality of 5G networks. Based on the proposed framework, we design a pipeline to exploit ROS-based information when creating and updating the NDT via a dedicated API.
% We make it public to foster research on related fields.\footnote{\new{Online available: \url{https://github.com/5G-ERA/NDT_5G_Robots_IROS24/}}}} 
%The integration of robotic data into the NDT for both network and robot control not only enhances network efficiency but also lays the foundation to enhance autonomous robotic systems. 
We explore the setup of an NDT at run time, including methods for data collection, modelling techniques, and their integration with robotic control algorithms. To validate the proposed NDT framework, we exploit robotic traces collected in an operational environment to showcase the evolution of an NDT model and its capabilities to adapt to variable network scenarios. And finally we show an example of using NDT for radio-aware navigation and highlight its benefits in improving energy efficiency and reliability for 5G robot operations.

%The main contributions of this work are summarized as follows:
%\begin{itemize}
%\item We propose a framework for the creation of an online NDT from a robot when exploring an unknown area for the radio quality estimation of the 5G networks.
%\item We design a pipeline to exploit ROS-based information when creating/ updating the NDT \new{ via a dedicated API. We make it public to foster research on related fields.~\footnote{\new{Online available: \url{https://github.com/5G-ERA/NDT_5G_Robots_IROS24/}} } }
%\item We exploit robotic traces collected in an operational environment to showcase the evolution of an NDT model and its capabilities to adapt to variable network scenarios.
%\end{itemize}

%The remainder of the paper is structured as follows.
%Sec.~\ref{sec:related} summarizes related works in the field.
%Sec.~\ref{sec:framework} describes the main architectural components and functionalities of the proposed solution, detailing a pipeline process for translating robot sensing data into quality of signal metrics that can be used by robots.
%Sec.~\ref{sec:perf_eval} validates the overall framework and evaluates its capabilities.
%Finally, Sec.~\ref{sec:conclusion} concludes this paper.

%%%%%%%%%%%%%%%%%%%%%%%%%%%%%%%%%%%%%%%%%%%%%%%%%%%%%%%%%%%%%%%%%%%%%%%%%%%%%%%%
\section{Related Works} 
\label{sec:related}

By leveraging the capabilities of NDT, digital replicas of networked environments can be created to offer unparalleled insights, predictive capabilities, and opportunities for optimization. 5G's low latency, high bandwidth, and massive device connectivity provide the foundation for real-time communication and control, enabling robots to readily respond to unexpected dynamics in the deployment environment. 

In this context, communication-aware motion planning in robotic networks is becoming of key importance in the robotic domain, due to the impelling need of offloading heavy computation tasks to edge platforms. 
The authors of~\cite{5GERA_IROS} investigate radio-aware semantic map creation along the exploration of unknown environments, and provide simulation-based results and a ROS package to deal with such scenarios. Similarly, \cite{online_OROS} investigate joint optimization of 5G and robotic domain as a way to improve resource allocation and energy consumption KPIs. %\cite{Delgado22} and
In the domain of wireless channel estimation within robotic networks, the authors of~\cite{5509677} propose a framework for estimating spatial variations in a wireless channel within a robotic network. They introduce a multiscale probabilistic model to characterize the wireless channel and develop an estimator based on this model.
In the area of NDT, most of the works in the literature consider holistic approaches by assuming full knowledge of the network infrastructure and environment, for example, the authors of~\cite{Graph_DTN} propose a knowledge graph-based construction method for NDTs, exploiting a graph representation to model the system dynamics. %In ~\cite{Scalable_DT} a scalable NDT for network slicing is developed, aiming to capture the intertwined relationships among slices and monitor the end-to-end (E2E) metrics of slices under diverse network environments. 
% \cite{next_generation_DT_robotics}
% \cite{SLAM_5GNR}
Altogether, this highlights the ongoing innovation and evolution in both the fields of robotics and communication to build an NDT, especially from a robot perspective, which needs to discover its environment and available network infrastructure. %, driven by the goal of meeting the demands of contemporary and future applications.
\new{Key challenges in building an NDT for robots stand on the model accuracy and real-time data acquisition. In this paper we introduce an online NDT model for robots exploring unknown areas, utilizing ROS-based real-time data to create and update NDT efficiently.}
% in terms of timeliness as well as the way to acquire and store data that the robots collect in real time. In this paper, we propose an online NDT model based on a robot when exploring an unknown area and develop a pipeline to exploit ROS-based real time information for creating and updating NDT over time.

%%%%%%%%%%%%%%%%%%%%%%%%%%%%%%%%%%%%%%%%%%%%%%%%%%%%%%%%%%%%%%%%%%%%%%%%%%%%%%%%
\subsection{Architecture Framework}
\begin{figure}[t]
\centering
\includegraphics[width=0.80\columnwidth, clip, trim = 4cm 0cm 3cm 1.2cm]{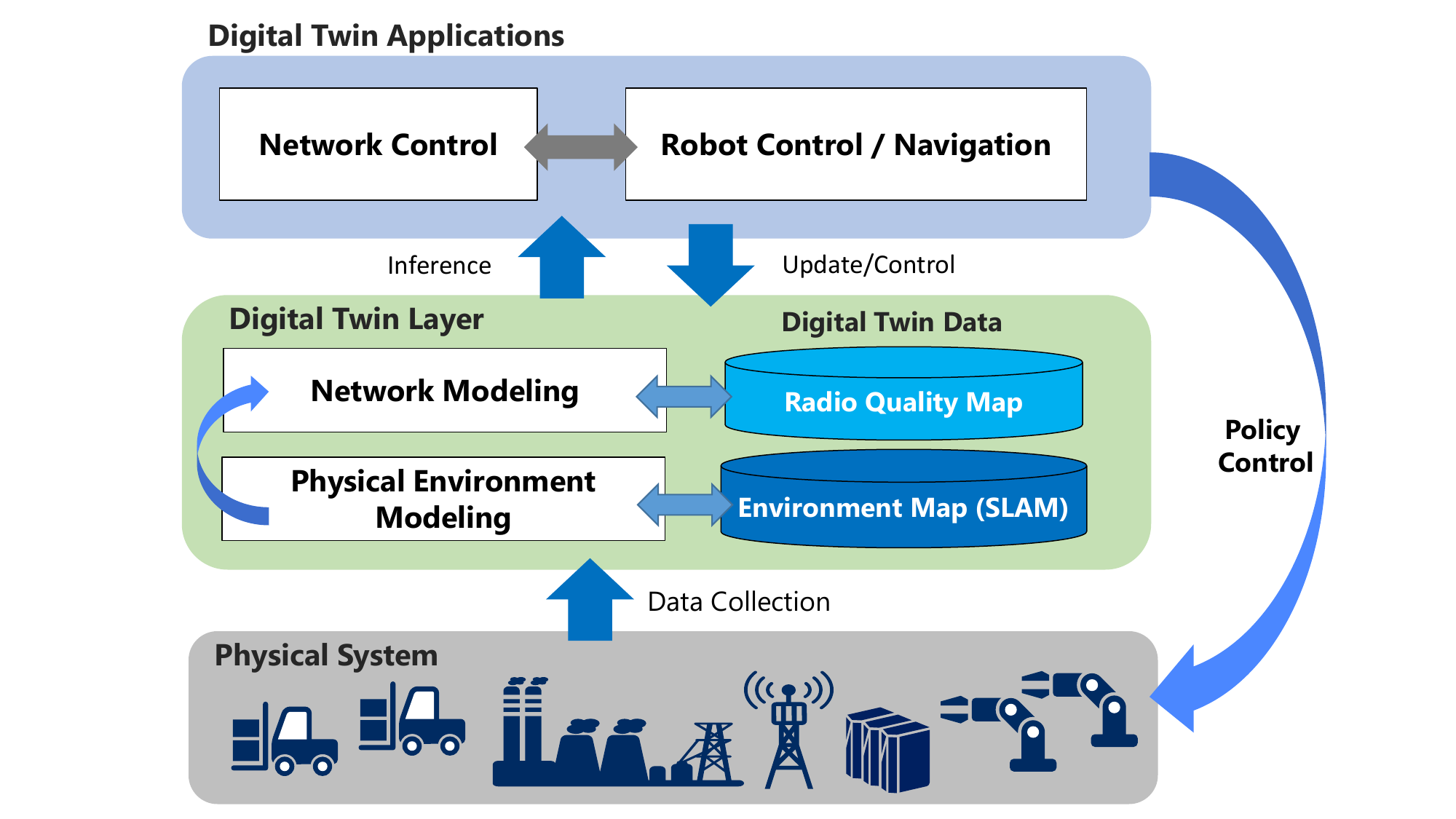}
\caption{Architecture of the Network Digital Twin for the joint network and robot control}
\label{fig:architecture}
 \vspace{-3mm}
\end{figure}

To build such an NDT to integrate with robotic data and applications and use it for joint network and robotic control, we design the architecture framework depicted in Fig.~\ref{fig:architecture}. It includes the Physical System, Digital Twin Layer, and Digital Twin Application layer. The Digital Twin Layer is the digital representation of the Physical System, which are interconnected with each other in real-time to realize the closed loop for the control of the real work system.  The digital representation is composed of Network Modeling and Physical Environment Modeling. The physical environment modeling creates the digital modelling of the physical environment, including the space, topology, objects, devices, humans, etc. 
The Network Modeling allows for building the virtual representation of the network behavior, as well as estimating the expected network performance for various applications including resource prediction, fault resilience, anomaly detection, and network control, together with robot control or navigation planning. 
The construction of the NDT models can be done by many different means such as using analytical models like queuing theory and network calculus, by means of simulations, or by adopting Machine Learning (ML)-based approaches, e.g., based on Graph Neural Networks (GNN), Deep Reinforcement Learning (DRL) or Deep Neural Network (DNN) models. Building the NDT models relies on the inputs from the collection of real data from various data sources from the Physical world system including monitoring data, system or operation data, or even raw data provided by the robots or UEs and different network and system entities. These data can be stored in the Digital Twin Layer for continuous updates and processing to facilitate the efficient usage of the time-series and large-scale data.  
 \vspace{2mm}
\section{Network Digital Twin Architecture}
\label{sec:framework}
\subsection{Network Digital Twin Model and Pipeline}
\label{subsec:pipeline}
% The control of mobile robots and data flow between the different functions and applications involved are aspects that have significantly progressed over the last few years.

\begin{figure}[t]
\vspace{4mm}
\centering
\includegraphics[width=.9\columnwidth, clip, trim = 0cm 3cm 0cm 2cm]{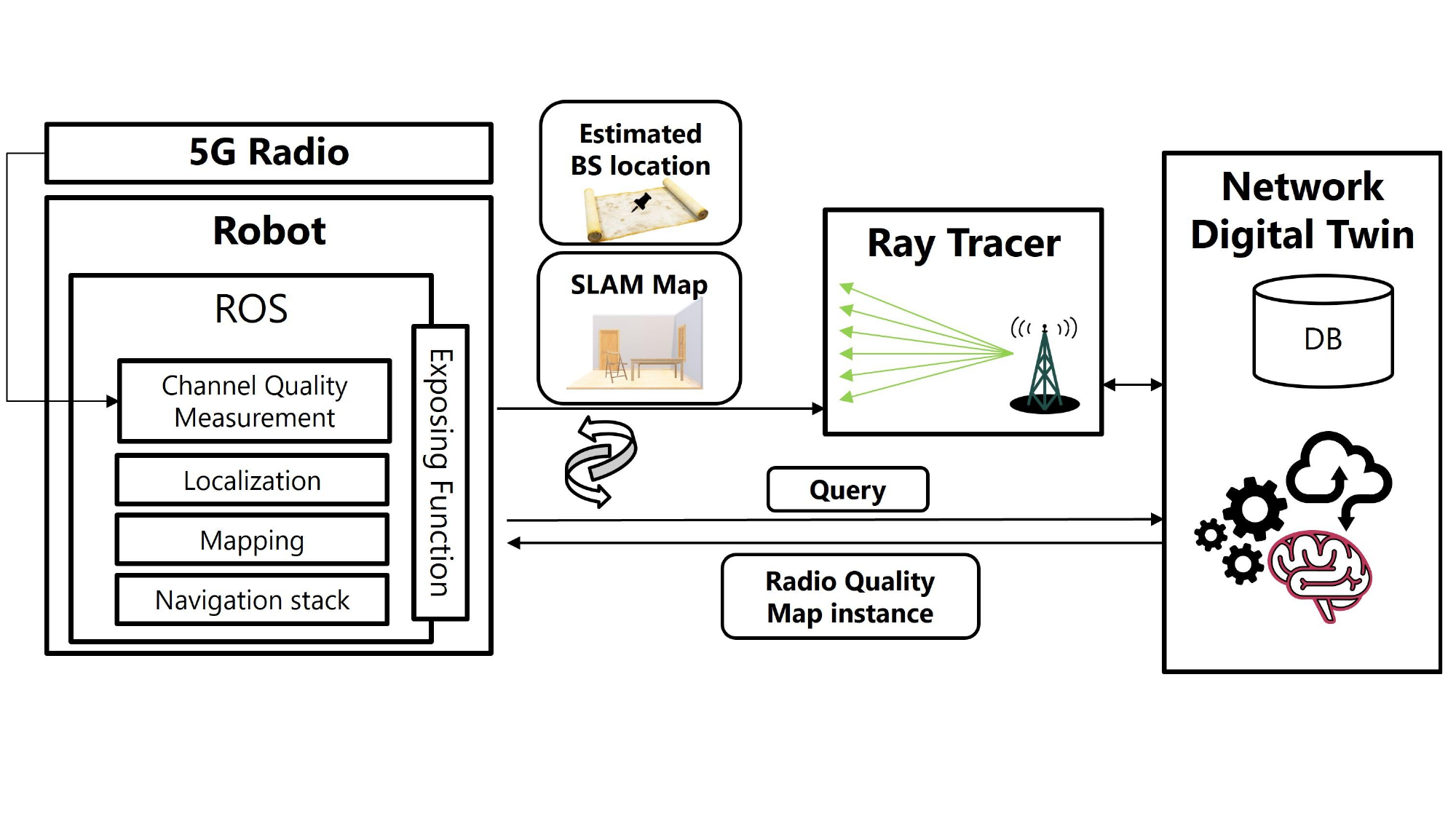}
\caption{Network Digital Twin Pipeline}
\label{fig:workflow}
\vspace{-3mm}
\end{figure}

Following the architecture framework depicted in Fig.~\ref{fig:architecture}, we design a pipeline for developing an NDT for optimizing the control of 5G-enabled mobile robots in a simulated 5G network environment, as illustrated in Fig.~\ref{fig:workflow}.\\
\textbf{Robot Operating System}. We rely on the ROS design and specifications for the setup and management of robotic applications as well as for both control and data plane communication aspects~\cite{ROS}. ROS is an open-source robotics middleware that provides common functionality (e.g., read sensor data, navigation, planning, etc.) over general hardware abstraction using low-level device control. It contains a collection of tools and libraries that simplify the setup of complex robotic applications across a variety of systems~\cite{Quigley2009}.\\
% ROS introduced the concept of \emph{topic} as a communication channel that allows different parts of a robotic system to exchange data.\\
\textbf{Simultaneous Localization and Mapping (SLAM)}. It is a fundamental function for autonomous robots. SLAM allows a robot to navigate an environment while simultaneously building a map of that environment and determining its own location within it. Various sensor data are involved in the SLAM process, such as laser scans, depth images, odometry and high-precision Inertial Measurement Unit (IMU) data. ROS makes use of specific topic messages to exchange information related to the SLAM process so that different software modules can work together seamlessly, thus allowing the robot to navigate and make decisions based on its understanding of the environment. Of particular importance in the context of our work is the presence of a Light Detection and Ranging (LiDAR) sensor. The LiDAR technology utilizes laser beams to measure distances and create 3D point cloud maps of the surroundings, enabling robots to perceive their environment accurately. Such information allows robots to navigate, avoid obstacles, and make informed decisions in real time, making it a powerful feature for robotics applications ranging from autonomous vehicles to industrial automation.\\
\textbf{Radio Network Simulator (Ray tracer)}.
Ray tracing software is employed extensively in 5G simulations for its capacity to accurately replicate the interactions of radio waves with objects in three-dimensional space. Within this context, the ray tracing process starts with the generation of virtual rays emanating from a simulated transmission source. These rays traverse the simulated environment, determining intersections with objects therein and calculating complex phenomena such as lighting, shadows, reflections, and refractions upon contact with solid objects in the environment. The collected data is then employed to render highly realistic images, making ray tracing indispensable for visualizing complex 3D scenes. In the realm of 5G simulations, ray tracing emerges as a pivotal technology due to its capacity to model accurate radio wave propagation, interference effects, and antenna performance. It enables network planners to optimize base station placement, assess indoor coverage, and enhance virtual reality and augmented reality applications by realistically rendering virtual objects and environments.

% \begin{figure}[t]
% \centering
% \includegraphics[width=\columnwidth, clip, trim = 0cm 0cm 0cm 0cm]{Figures/3.final_wo.png}
% \caption{Example Voxel Image obtained from the 3D pointcloud}
% \label{fig:voxel}
% \end{figure}

\subsection{Base Station Position Estimation}
An accurate estimation of the radio environment demands knowledge about the location of the serving base station on the robot side. While multiple schemes are available in the literature focusing on how to locate a UE in the network (especially in the context of MIMO and beam forming, e.g., ~\cite{5GPositioning_Survey}), the opposite is less common as it generally implies heavy data collection and processing that may affect the performances of battery constrained devices like mobile phones.
The most common schemes in the literature can be split into two main categories: \emph{i}) approaches based on Received Signal Strength (RSS) and \emph{ii}) approaches based on the geometric and spatial distribution of measurements.
The first approaches have shown significant advantages in radio-channel fingerprinting~\cite{smallcell_discovery}, especially when adopted in conjunction with Machine-Learning schemes~\cite{NN_Assisted_localization}. These approaches, however, rely on an accurate characterization of the communication path loss, which may result difficult in urban environments characterized by Non-Line-of-Sight (NLOS) communication and numerous reflections.
On the other side, geometric methods rely on the spatio-temporal distribution of the measured metrics, and on deployment-specific information such as BS sectorization to enhance their estimations~\cite{ICRA24}. For example, 5G provides a dedicated Downlink (DL) signal for UE localization through the Positioning Reference Signal (PRS)~\cite{38.211}. An accurate UE position estimation can then be derived through the time difference of arrival of PRSs from multiple BSs via multilateration~\cite{Book_Positioning_LTE}.
In the context of our work, we rely on the Timing Advance (TA) measurements as described in~\cite{localizing_bs} for BS location estimation. More into details, the TA is an integer value $n \in \mathcal{N}, \mathcal{N}=\{0,\dots, N-1\}$ that the serving BS uses to maintain synchronization among the  connected UEs, ensuring that signals transmitted by multiple UEs arrive at the BS at the right time, without interference or collision. In 3GPP standards, N-1 takes a maximum value of 3486~~\cite{TS138213}. In practical terms, and assuming propagation with the speed of light in the vacuum $v$, the round trip time between the BS and the UE can be defined as a discrete set $d_n \in \mathcal{D} = \{d_0, \dots, d_{N-1}\}$, and estimated by:
    $d^{TA}_n = N^{TA}_n \cdot T_{c} \cdot v$,
% \begin{equation}
%    d^{TA}_n = N^{TA}_n \cdot T_{c} \cdot v \qquad [m]
%\end{equation}
where $N^{TA}_n = n\cdot 16\cdot64\cdot2^{-\mu}$ is the time duration expressed in 5G New Radio time-slots associated to the \emph{n}-th timing offset~\cite{TS138213} (Sec. 4.2), $\mu$ is the 5G numerology adopted, and $T_{c} = \frac{1}{\Delta f_{max} \cdot N_f}$ is the time 5G New Radio time slot duration, assuming $\Delta f_{max}= 480\cdot10^3$ Hz and $N_f = 4096$ as defined in~\cite{38.211} (Sec. 4). The resulting granularity for different 5G New Radio numerology settings is summarized in Table~\ref{tab:5GNR}.

While such kind of estimation would require significant energy consumption in UE devices, robots are generally equipped with larger batteries, making such processing less impactful on the overall consumption.

%\cite{LTE_positioning}

%%%%%%%%%%%%%%%%%%%%%%%%%%%%%%%%%%%%%%%%%%%%%%%%%%%%%%%%%%%%%%%%%%%%%%%%%%%%%%%%
\section{Performance Evaluation}
\label{sec:perf_eval}

\subsection{SUMMIT-XL 5G-Enabled Robot}
\label{subsec:robot}
\begin{figure}[t]
\centering
\vspace{4mm}
\includegraphics[width=.7\columnwidth, clip, trim = 2cm 0cm 1cm 0cm]{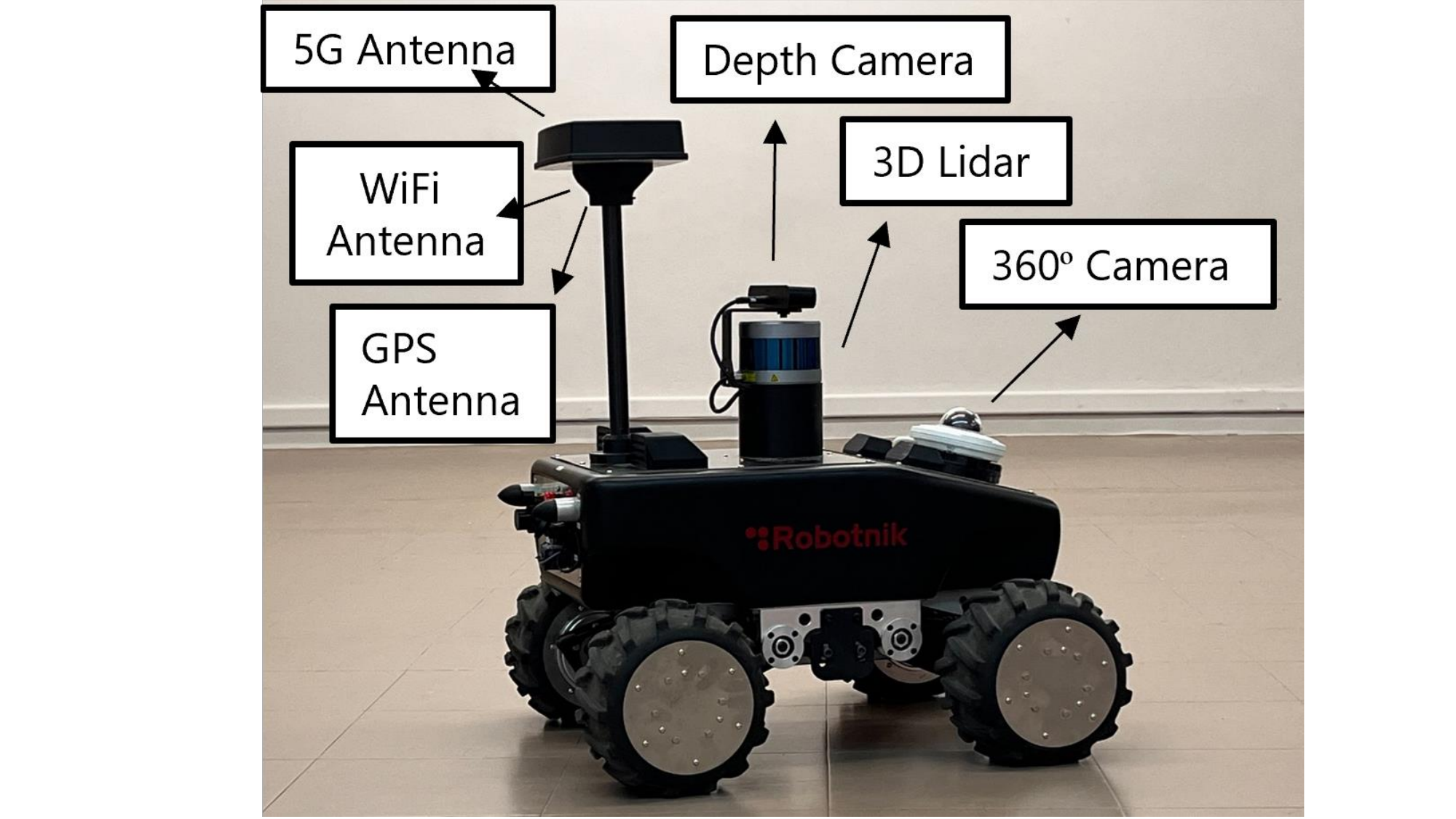}
\caption{SUMMIT-XL Robot}
\label{fig:summitXL}
\vspace{-1mm}
\end{figure}

In order to implement and evaluate our proposal, we rely on real traces collected by the Robotnik's SUMMIT-XL robot~\cite{SUMMIT-XL} depicted in Fig.~\ref{fig:summitXL}. This autonomous mobile robot can carry loads up to $50$ kg, and it is equipped with a $500$ W brush-less motor mounted on each rubber wheel, providing high mobility in unstructured environments. For communication, it relies on a 5G modem (Teltonika RUTX50) and external antennas. In order to sense the environment, this unit is equipped with a high-precision IMU, an HD Camera as well as a stereo depth camera. On the computing side, the robot is equipped with a $10^{th}$ generation Octa-core Intel i7-10700 and an additional ARM Nvidia Jetson GPU. It runs Ubuntu Linux $22.04$ and hosts the robotic core software using ROS2 Humble and related functionalities as containers. More interesting for our work, the robot is also equipped with an RS-LIDAR-16 3D LiDAR from RoboSense. This device operates at $905nm$ and allows collecting LiDAR samples over a $360^\circ$ horizontal field of view with a tunable resolution of $0.1^\circ,0.2^\circ,0.4^\circ$. The vertical field of view spans $30^\circ$ with a resolution of $2^\circ$. The device rotation speed reaches $300/600/1200$ rpm, which translates into a frame rate of $5/10/20$ Hz, respectively. The device generates $3\cdot10^6$ points/s, with an accuracy of $\pm2$ cm.

\begin{table}[b]
    \centering
    \caption{5G TA localization granularity $d_{TA}$ in meters for different numerology (and Subcarrier spacing) configuration $\mu$.}
    \label{tab:5GNR}
    \renewcommand{\arraystretch}{1.1}
    \begin{tabular}{c||c|c|c|c|c}
        & $\mu=0$ & $\mu=1$ & $\mu=2$ & $\mu=3$ & $\mu=4$ \\
        \hline
        $d^{TA}$ [m] & 78.125 & 39.063 & 19.531 & 9.766 & 4.883 \\
    \end{tabular}
\end{table}

\subsection{Scenario Setup}
\label{subsec:scenario}

\begin{figure}[t]
\centering
\includegraphics[width=.89\columnwidth, clip, trim = 6cm 3cm 6cm 2.5cm]{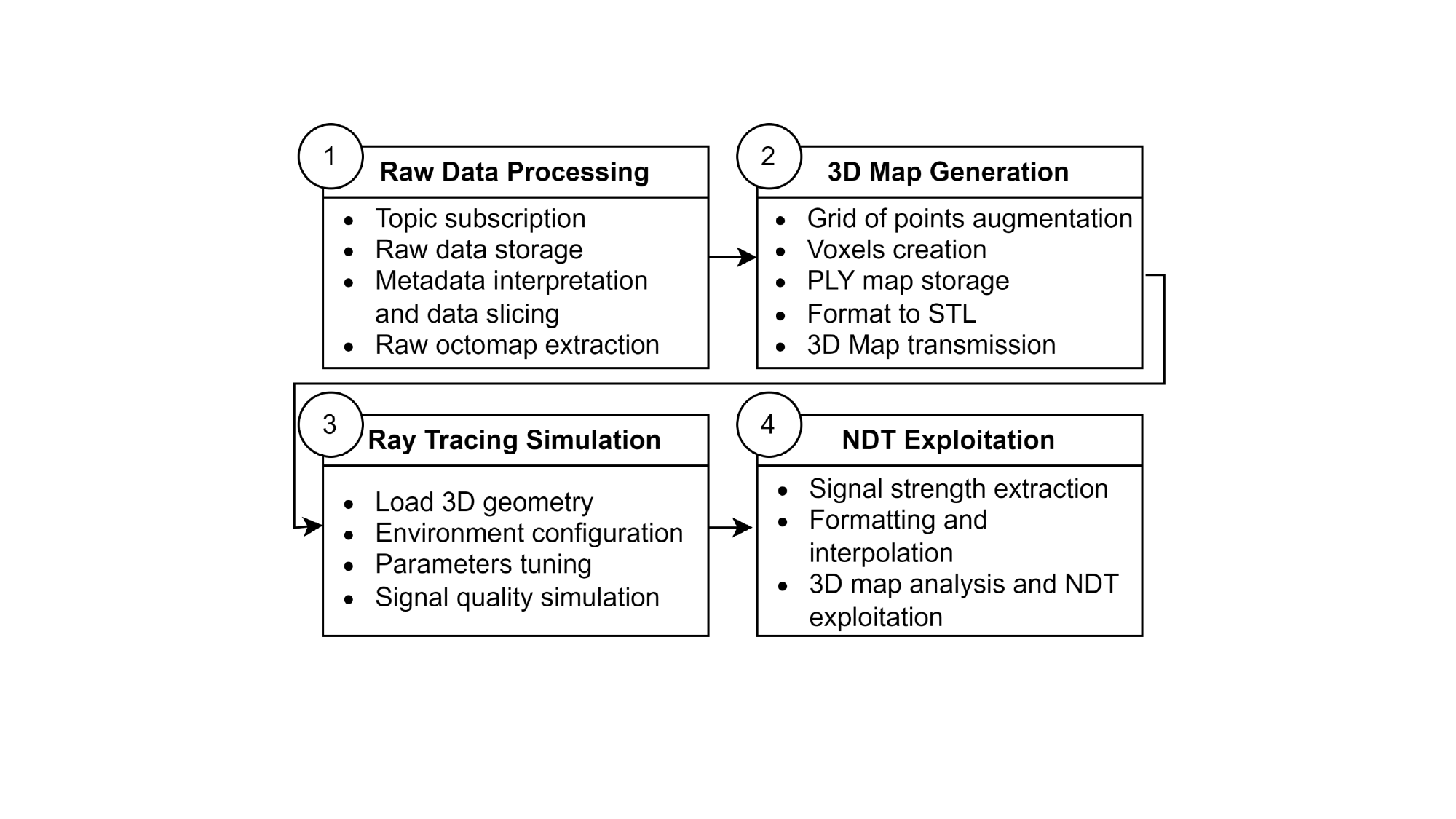}
\caption{Stages to extract, process and transform raw data into a 3D radio quality map.}
\vspace{-1mm}
\label{fig:data_stages}
\end{figure}

%After discussing in the previous section the essential tools required for communication and mapping of the robot parameters to achieve localization within the environment, it is necessary to describe in detail the implementation of the solution on top of such tools in order to enable the construction of a coherent 3D map from LiDAR sensor data.
Point cloud and SLAM data, obtained from various sensors like LiDAR and RGB-D camera are essential data sources in robotics and computer vision applications. Converting such data into a voxel representation is crucial for efficient spatial analysis. 
For this purpose, we leverage ROS modular architecture and 
create a dedicated package capable of subscribing to Point2Cloud messages, processing the data, and publishing the resultant 3D pointcloud information. The process involves conversion from ROS message format to Point Cloud Library (PCL) data structures.
The resulting map is stored in a database, and subsequently used by the ray tracer to estimate the signal quality in each point of the environment. To this aim, we design the 4-staged pipeline depicted in Fig.~\ref{fig:data_stages}.

\begin{figure}[t]
\centering
\subfloat[LiDAR data]{\includegraphics[width=0.45\columnwidth, clip, trim = 7cm 0cm 9cm 0cm]{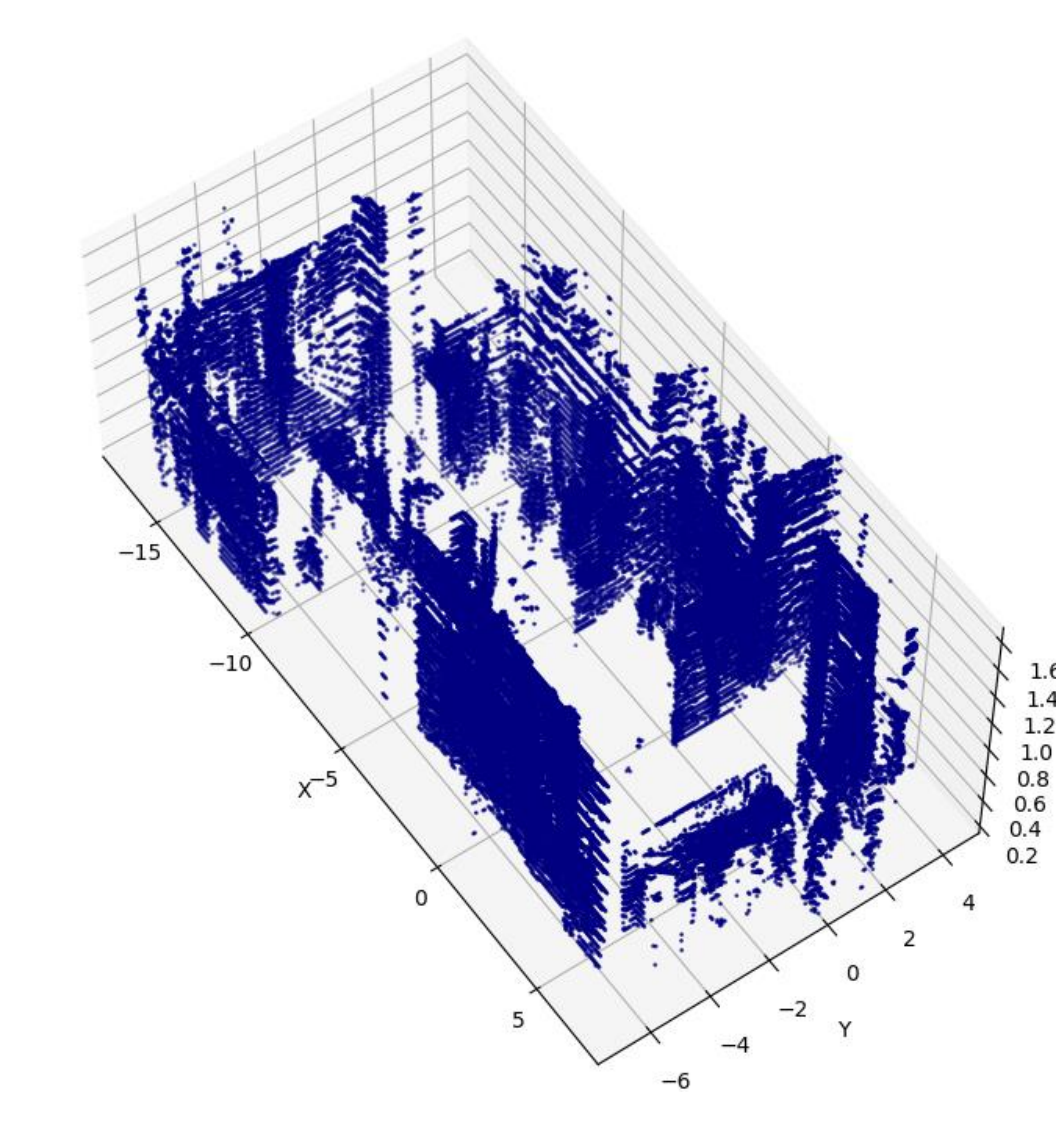}%
\label{fig:pointcloud}}
\hfil
\subfloat[Voxels 3D map]{\includegraphics[width=0.47\columnwidth, clip, trim = 6cm 0cm 6cm 0cm]{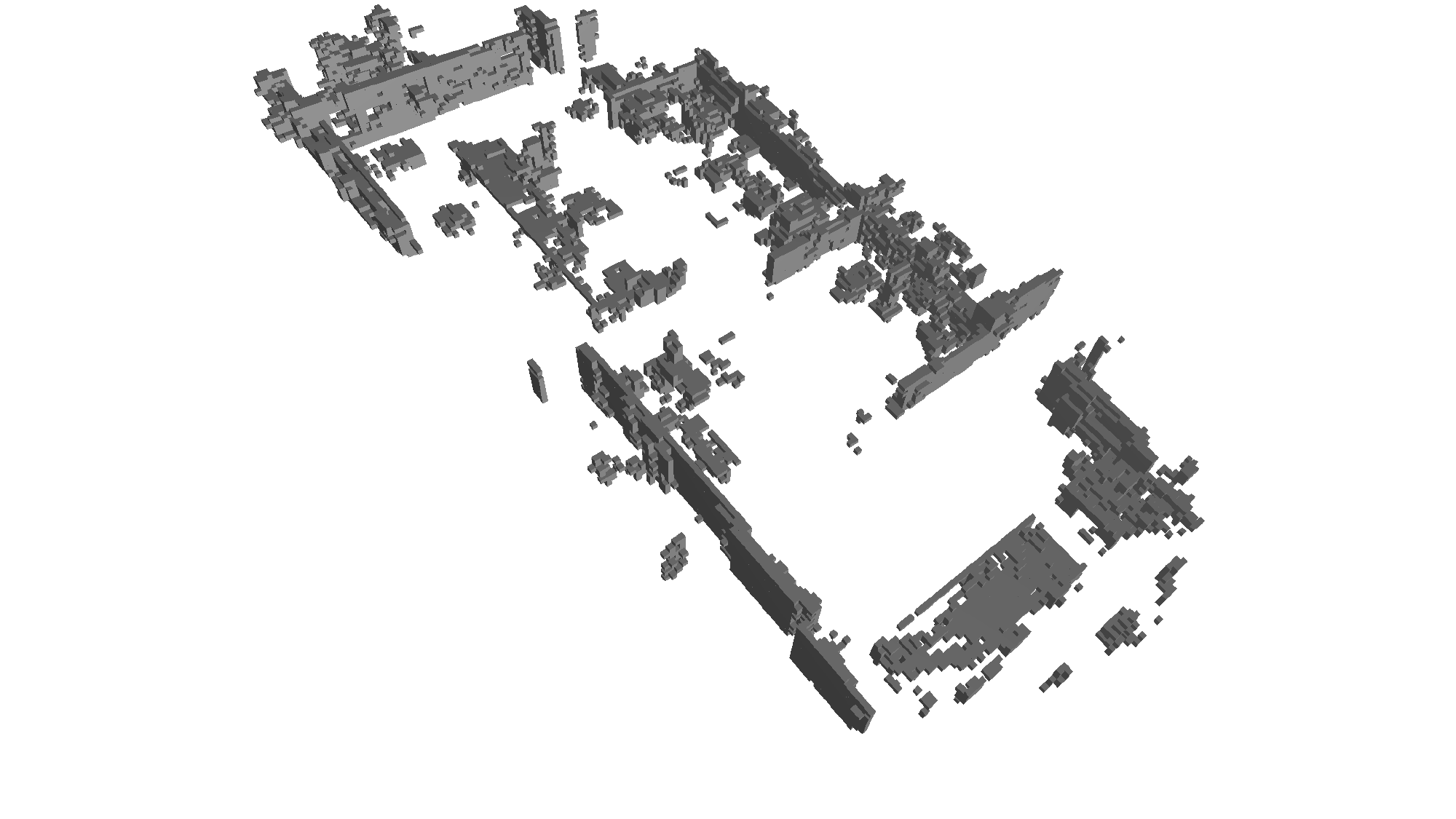}%
\label{fig:voxel}}
\hfil
\caption{Pointcloud data measured from the 3D LiDAR sensor (left) and resulting 3D voxels (right)}
\label{fig:pointcloud_voxel_comparison}
\vspace{-3mm}
\end{figure}

\textbf{Raw Data Processing} The data coming from the 3D LiDAR sensor is merged with the information provided by the SLAM algorithm to build our 3D map taking into account the relative position of the robot with respect to the room and persistent objects inside it. A representation of the raw data can be found in Fig.~\ref{fig:pointcloud}.
To this end, we exploit the OctoMap library
%~\cite{octomap}
to perform the three-dimensional and occupancy grid mapping considering the SLAM information. Octomap uses 3D pixels to represent the environment, where each point in space is associated with a probability value indicating its occupation status.
More into detail, the ROS topic containing the collected LiDAR information is organized in a YAML format as follows:
\begin{itemize}
    \item \textit{height} and \textit{width}: indicate the dimensions of the data.
    \item \textit{datatype}: determines the datatype and format of the data. In our case, the value is fixed to seven, corresponding to the \textit{Float32} datatype.
    \item \textit{is\_bigendian}: indicates whether the data is encoded using the big-endian format. In our case, it is set to \textit{False}, meaning the data is stored in little-endian format.
    \item \textit{point\_step}: specifies the size of each point. In our settings, its value is fixed to 16 bytes, i.e., four bytes for each of the four fields, X, Y, Z, and intensity, respectively.
\end{itemize}

\begin{figure*}[ht!]
\centering
\subfloat[Initial 3D map]{\includegraphics[width=0.24\textwidth, clip, trim = 7cm 4cm 7cm 3cm]{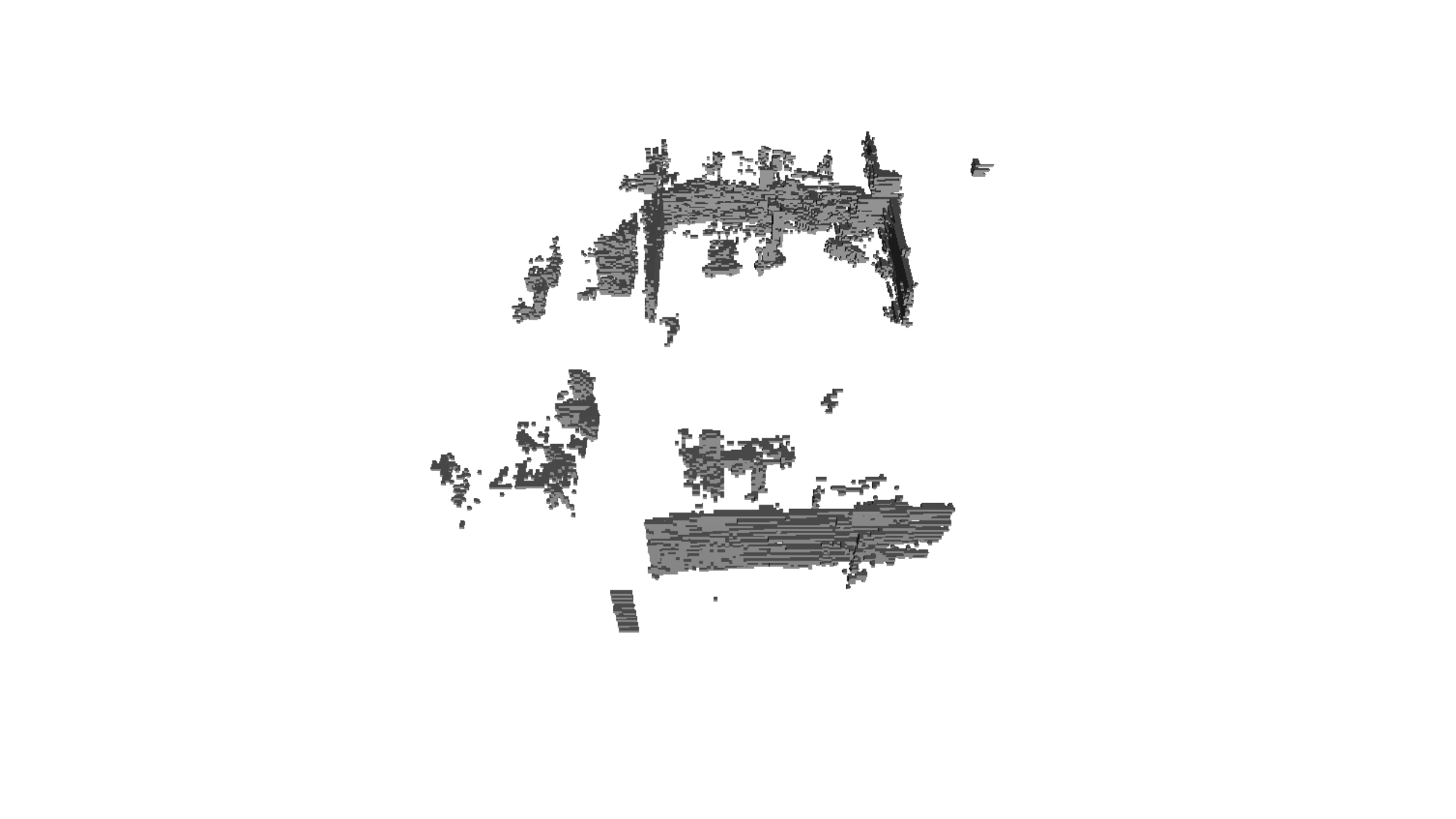}%
\label{fig:map_evolution_1}}
\hfil
% \centering
\subfloat[Intermediate 3D map]{\includegraphics[width=0.22\textwidth, clip, trim = 7cm 3cm 9cm 3.5cm]{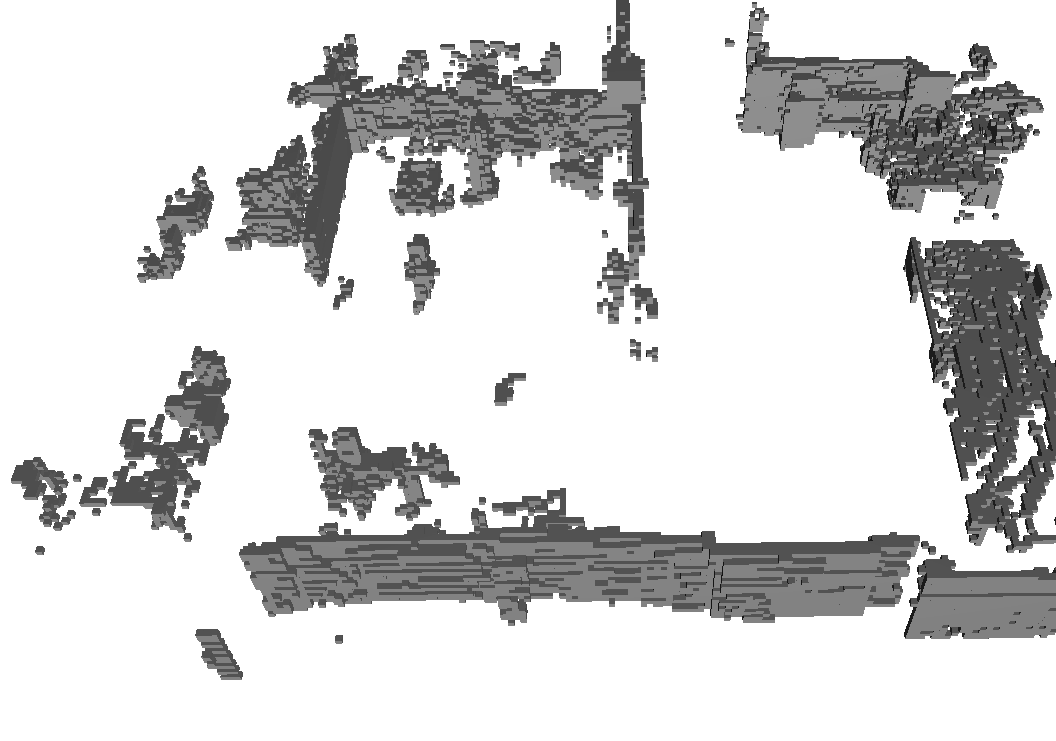}%
\label{fig:map_evolution_2}}
\hfil
% \centering
\subfloat[Final 3D map]{\includegraphics[width=0.43\textwidth, clip, trim = 0cm 4.5cm 0cm 3.5cm]{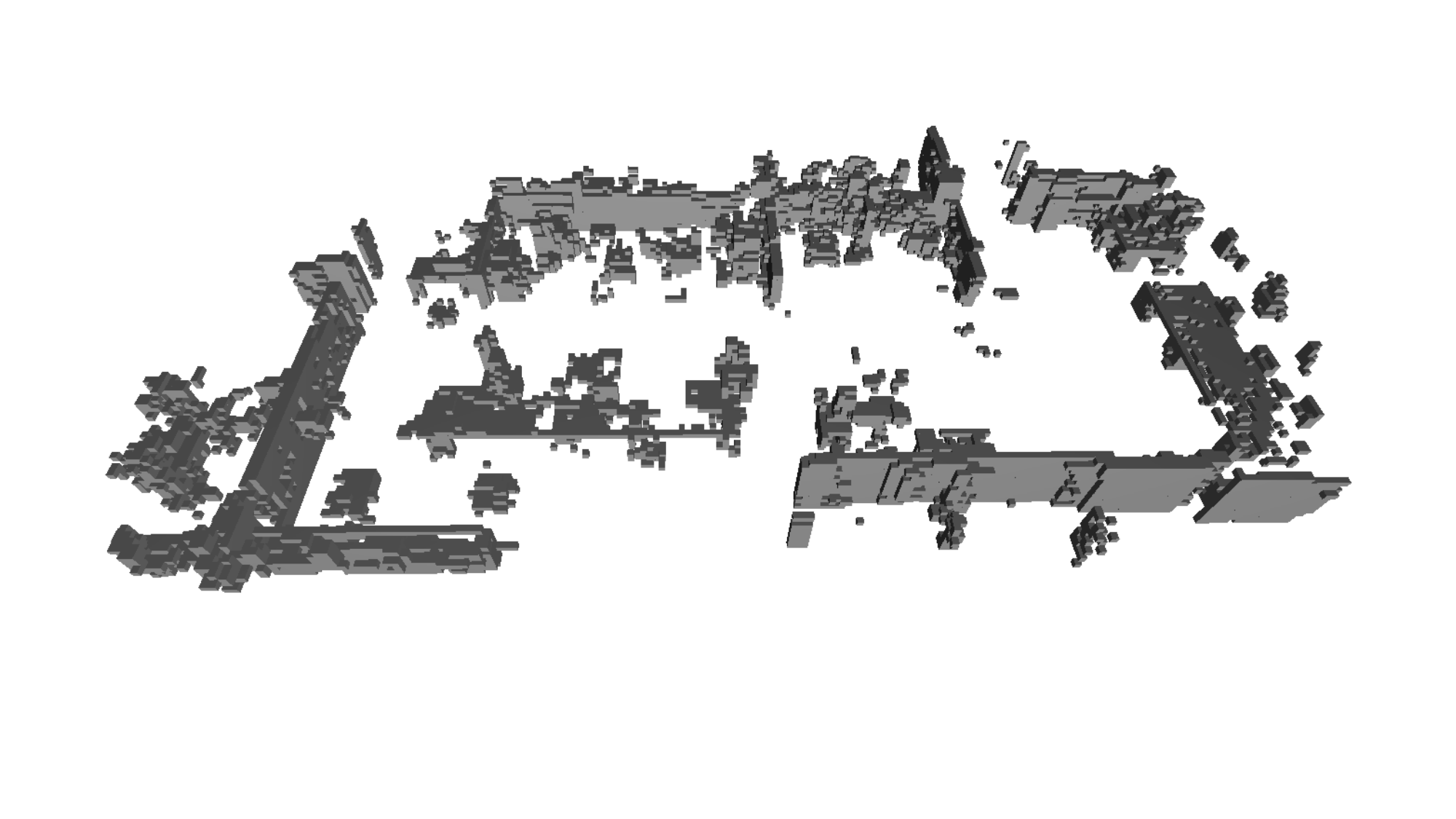}%
\label{fig:map_evolution_3}}
\hfil
\caption{3D map evolution collected by merging LiDAR data during the SLAM process}
\vspace{-3mm}
\label{fig:SLAM_map_evolution}
\end{figure*}

\begin{figure*}[ht!]
\centering
\subfloat[Initial radio quality map]{\includegraphics[width=0.33\textwidth]{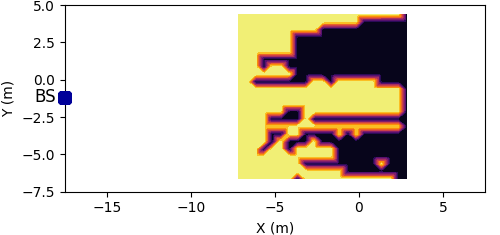}%
\label{fig:ndt_evolution_1}}
\hfil
\subfloat[Intermediate radio quality map]{\includegraphics[width=0.33\textwidth]{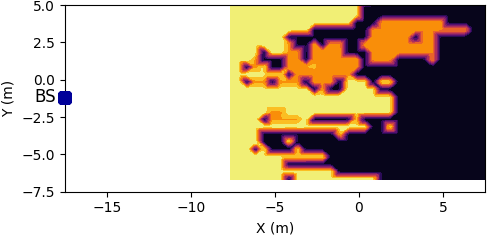}%
\label{fig:ndt_evolution_2}}
\hfil
\subfloat[Final radio quality map]{\includegraphics[width=0.34\textwidth]{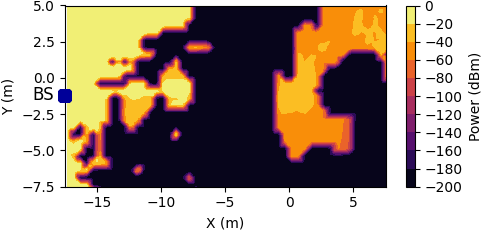}%
\label{fig:ndt_evolution_3}}
\vspace{-4mm}
\subfloat[Initial radio quality map]{\includegraphics[width=0.33\textwidth]{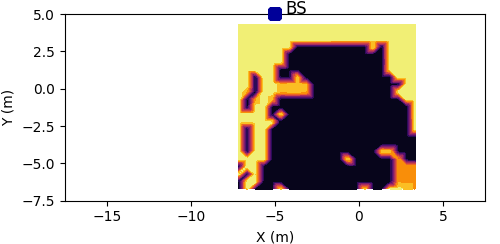}%
\label{fig:ndt_north_1}}
\hfil
\subfloat[Intermediate radio quality map]{\includegraphics[width=0.33\textwidth]{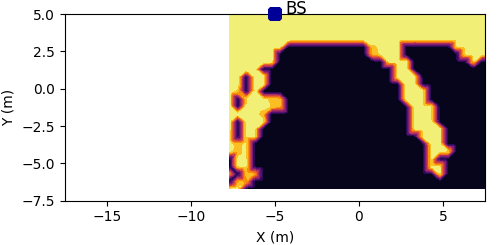}%
\label{fig:ndt_north_2}}
\hfil
% \subfloat[Final radio quality map]{\includegraphics[width=0.34\textwidth]{Figures/3.North_Brick_Final.png}%
\subfloat[Final radio quality map]{\includegraphics[width=0.32\textwidth]{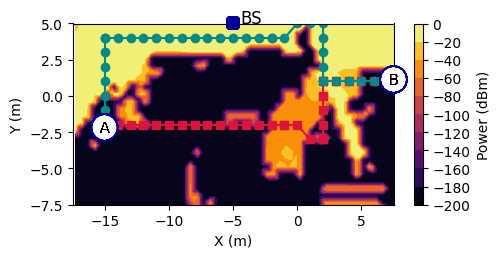}%
\label{fig:ndt_north_3}}
\hfil
\caption{Radio quality map evolution in different scenarios: 1) BS located on the north side (upper row) and 2) BS located on the west side (lower row).}
\label{fig:radio_evolution_north}
% \label{fig:radio_evolution_west}
\vspace{-3mm}
\end{figure*}

\textbf{3D Map Generation} Once the data are extracted and processed, an augmentation is performed in order to fill the gaps between objects with virtual data obtained by interpolation. This leads to a 3D map consisting of estimated and real points in space. Then, we make use of an open-source library for 3D data processing~\cite{open3d} to derive a 3D voxel (short for "volume elements") grid in Polygon File Format (PLY). The resulting PLY file is converted to STereoLithography (STL) format, the expected format in our ray tracer software. Fig.~\ref{fig:pointcloud_voxel_comparison} depicts an example 3D pointcloud data collected through LiDAR sensor, and the resulting voxel representation.

\textbf{Ray Tracing Simulation} The third stage focuses on estimating the radio propagation settings in the virtual 3D environment. To accurately recreate the radio propagation environment, we adopt the Wireless InSite software~\cite{InSite} a 3D ray-tracing simulator widely used in the research community to analyze site-specific radio wave propagation and wireless communication systems. 
We chose the X3D Ray Model to achieve high accuracy through the exact path calculations.
% The chosen propagation model is the X3D Ray Model, a full 3D propagation model with a high accuracy achieved through the exact path calculations.
The computing platform hosting the ray tracer is equipped with two Intel Xeon CPU@2.10 GHz with 32 cores and 128 GB of RAM. We consider a single RAN node configured with numerology $\mu=4$ (i.e., subcarrier spacing $240$KHz), transmission power of 43~dBm~\cite{TS38.104} at 1800~MHz, i.e., band 3, and set a grid of receivers at $1$m height to emulate the antenna height of the SUMMIT-XL while operating. Notably, within these settings, the communication may be influenced by static objects, e.g., office furniture, typically present in indoor environments.
As the map size grows following the robot exploration, the receiver grid size is adjusted to capture the signal emitted by the BS in every location of the map. Without loss of generality, we fix the interspacing of the receivers to $0.5$m. Such a parameter may be easily reduced to improve spatial granularity, at the cost of larger computation requirements.\\
\indent \textbf{NDT Exploitation} The simulation results are used in the final stage to extract the signal strength captured by the receivers. We perform an interpolation to estimate the power of the received radio signal at every point in the environment. \new{Traditional exploration schemes measure and aggregate local information about the radio characteristics by navigating in every section of the environment~\cite{5GERA_IROS}. Conversely, our approach allows the robot to gain knowledge 
of the expected radio quality for a given environment within one interaction with the remote service}. The obtained signal quality map can be used as the baseline of the NDT to exploit different robot control and navigation configurations before applying them to the robots in the physical world. For example, the NDT could provide the expected radio information along a planned trajectory, allowing the robot to adapt its behaviour accordingly.

\subsection{Evaluation Results}
\label{subsec:results}
The model capabilities to match the real environment characteristics directly depend on the quality of the collected data and the reconstructed 3D map. In order to show the evolution of both environmental and radio quality maps along our experiment, we select three snapshots of a data collection experiment taking about 10 minutes in total.

\begin{figure}[t]
\centering
\includegraphics[width=.65\columnwidth, clip, trim = 5cm 3cm 6cm 3cm]{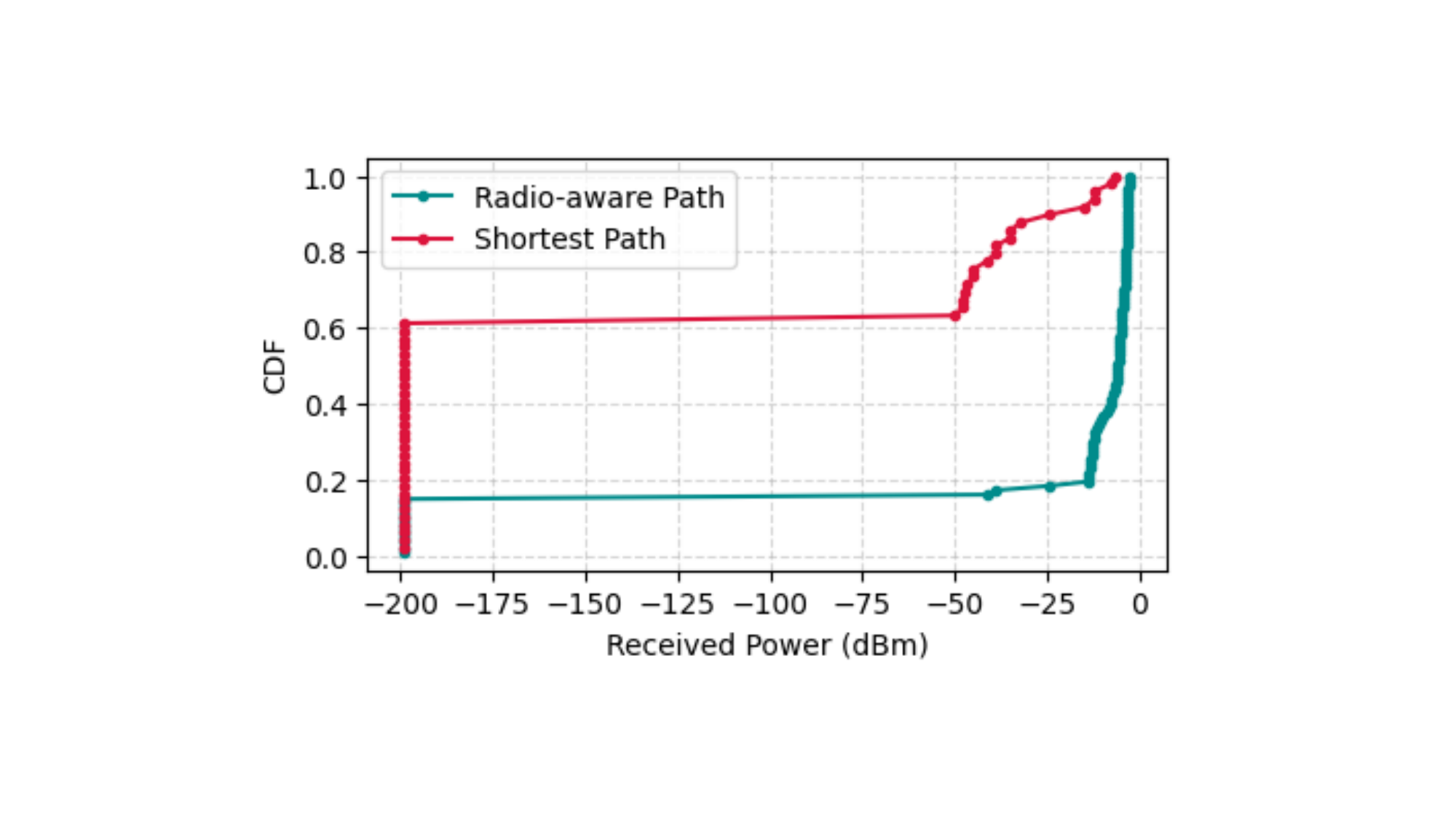}
\vspace{-1mm}
\caption{Comparison of the received signal power for two different trajectories.}
\label{fig:trajectories}
\vspace{-2mm}
\end{figure}

Fig.~\ref{fig:SLAM_map_evolution} shows the evolution of the 3D map collected by the robot, which is used as input geometry for the ray tracer.
Within this scenario, we place the BS $10$ m away from the test area with a fixed height of two meters, specifically located at $(x,y,z) = (-27.5,-1,2)$ meters.
In particular, Fig.~\ref{fig:map_evolution_1} depicts the collected environmental map after about one minute. Comparing it with Fig.~\ref{fig:ndt_evolution_1}, it can be noticed how the resulting radio quality map tends to overestimate the received radio power. This is due to multiple walls and objects present in the environment have not been detected by the robot, which translates into a free space propagation scenario from the ray tracer perspective.
Nevertheless, the knowledge about the real environment increases together with the robot exploration phase, allowing for a more accurate estimation of the radio environment characteristics, as it can be appreciated by comparing Fig.~\ref{fig:map_evolution_2} and Fig.~\ref{fig:ndt_evolution_2}, which represents the status of the system at about three minutes from the beginning of the experiment. The map increases in size and becomes progressively more detailed in the representation of objects.
Finally, Fig.~\ref{fig:map_evolution_3} and Fig.~\ref{fig:ndt_evolution_3} depict the final phase of the experiment, where the robot has been able to collect all the details of the 3D environment and the corresponding radio power information. Notably, most of the signal power enters the room by a window located on the top left side of the map (in Fig.~\ref{fig:map_evolution_3}). Such information would be essential when planning robot navigation to ensure adequate radio coverage for, e.g., offloading purposes. The NDT is constantly enriched with real-time data, allowing for improved accuracy in the estimation of the radio quality.
Moreover, the capabilities of the NDT model to meet real-world dynamics also strictly depend on an accurate BS localization, given that the radio propagation is affected by the geometry of the environment and its physical properties. Such information is assumed to be an input of our solution.
To showcase the effects of an erroneous BS localization, we performed another experiment maintaining the same $10$ m distance but changing its location towards the north side of the map, i.e., placing the transmitting antenna in $(x,y,z) = (-5,14,2)$ meters.
Results are depicted in Fig.~\ref{fig:radio_evolution_north}, which summarizes the evolution of signal strength captured by the receivers in the ray tracer.
We can clearly appreciate how the received signal strength radically changes, and how the areas with better signal strength have been shifted from the central left area to the right side. This highlights the importance of knowing the location of the BS for efficient task planning, either by optimizing the route to find paths with reliable signals~\cite{5GERA_IROS}, or by enabling offloading only in areas where the robot expects good coverage~\cite{online_OROS}, or a combination of the two. Such radio-aware navigation may positively improve both energy savings and communication reliability.
% Last addition 
To showcase the inherent potential benefits of our approach, we simulate two indoor trajectories depicted in Fig.~\ref{fig:ndt_north_3} (from point A to point B) considering radio-aware navigation and a simple shortest path algorithm. Fig.~\ref{fig:trajectories} compares the CDF of the received signal power collected along the navigation paths. From the plot, we can notice how radio-aware navigation provides stronger radio connectivity when compared to standard shortest-path navigation algorithms, finally ensuring more reliable teleoperations and computation offloading capabilities for the robots along its movement.
%
% \vspace{-2mm}

%%%%%%%%%%%%%%%%%%%%%%%%%%%%%%%%%%%%%%%%%%%%%%%%%%%%%%%%%%%%%%%%%%%%%%%%%%%%%%%%
\subsection{Discussions}
\label{sec:discussion}

% The localization accuracy in 5G systems is significantly improved compared to LTE, but limitations still exist due to inherent challenges. Timing Advance (TA) in 5G operates with finer subcarrier spacing and smaller timing granularity than LTE, enabling more precise distance resolution. However, factors such as multipath propagation, non-line-of-sight (NLoS) conditions, and interference from dense deployments of base stations can still impact accuracy, particularly in urban and indoor environments where reflections and obstructions are prevalent. Additionally, the resolution is limited by the timing synchronization and the geometry of the network, where poor base station placement or insufficient anchor points can degrade localization. To overcome these challenges, 5G leverages ultra-wideband signals, higher carrier frequencies, and advanced beamforming techniques, which allow for more precise angle-of-arrival (AoA) and time-of-flight (ToF) measurements. The introduction of mmWave frequencies provides finer spatial resolution, while multi-base station coordination and enhanced positioning protocols, such as enhanced Observed Time Difference of Arrival (eOTDoA), improve accuracy even in challenging conditions.

The localization accuracy in 5G systems exploits finer Timing Advance (TA) granularity, ultra-wideband signals, and higher frequencies like mmWave, enabling sub-meter precision.
However, challenges like multipath propagation, NLoS conditions, and interference in dense deployments can still degrade accuracy. Solutions include enhanced Observed Time Difference of Arrival (eOTDoA), multi-base station coordination, beamforming, and AI-driven multipath mitigation providing robust positioning even in complex environments. %
Moreover, the computation time required to update the signal information in the NDT model depends on multiple factors, including the size of the environment, the desired spatial granularity of the radio information, and the computing platform itself. \new{Due to the complexity of the ray-tracing process and the concurrent lack of GPU equipment, the computation time in our test scenario can take up to several minutes, therefore failing to meet the real-time requirements of some robotic applications.} 
In this regard, ML-based approaches can be adopted to speed up the computation and achieve more efficient models. The proposed NDT framework and pipeline have built the foundation for developing ML-based solutions that allow real-time digital twin computation of various ``what-if" scenarios to optimize robot decision-making, control and navigation, thus empowering robot autonomy.
%Furthermore, the current results are based on a static scenario. However, the radio propagation environment of real-world systems may be affected by moving objects in the environment. How to keep track of or predict such variations in the NDT model is still an open research topic.

%close to half an hour in the case of the initial map, increasing by 15 min for the intermediate map and close to an hour if we simulate the more complex map. %Consequently, in the worst scenario, we could upgrade our NDT every hour with an enhanced radio quality map. 

%%%%%%%%%%%%%%%%%%%%%%%%%%%%%%%%%%%%%%%%%%%%%%%%%%%%%%%%%%%%%%%%%%%%%%%%%%%%%%%%
\section{Conclusions and future work}
\label{sec:conclusion}

The convergence of NDT technology with the deployment of 5G-enabled robots represents a significant leap forward in the realm of wireless communication, automation, and intelligent systems.
In this paper, we develop a framework and pipeline for the implementation of an online NDT approach in the context of 5G-enabled robots, focusing on the exploration of unknown areas and estimation of the expected radio quality of discovered areas at run time. We rely on real LiDAR traces to derive the 3D environmental map and simulate realistic 5G settings to validate our proposal. Our results show that NDT technology leads to significant improvement of robot operations by gaining knowledge of estimated radio coverage and signal quality, making it a promising solution for the future of connected 5G-enabled robot systems. In future work, we plan to explore ML-based models to further enhance the NDT real-time capabilities and improve its computation efficiency for highly dynamic environments.

%%%%%%%%%%%%%%%%%%%%%%%%%%%%%%%%%%%%%%%%%%%%%%%%%%%%%%%%%%%%%%%%%%%%%%%%%%%%%%%%

\section*{Acknowledgment}
The research leading to these results has been supported by the European Union's H$2020$ 6GGOALS Project (grant no. $101139232$), H$2020$ MultiX Project (grant no. $101192521$)  and by the UNICO I+D 5G 2021 number TSI-063000-2021-14/15.

% MINECO 6GSmart project TSI-063000-2021-14/15. 

% The research leading to these results has been supported by the European Union's H$2020$ 6GGOALS Project (grant no. $101139232$). 
% We further acknowledge the Grant DIN2018-010177 and PID2021-123627OB-C52 funded by MCIN/AEI/10.13039/501100011033 and by ``ERDF A way to make Europe''.

%%%%%%%%%%%%%%%%%%%%%%%%%%%%%%%%%%%%%%%%%%%%%%%%%%%%%%%%%%%%%%%%%%%%%%%%%%%%%%%%
\bibliographystyle{IEEEtran}
\bibliography{main}
\end{document}